\documentclass[12pt]{article}
\usepackage{amsmath}
\usepackage{amsfonts}
\usepackage{amssymb}
\usepackage{graphicx}
\usepackage{bm}
\usepackage{hyperref}
\usepackage{color}
\usepackage{mathrsfs}
\usepackage{epsf}

\begin{document}
\newcommand{\la}{\label}
\newcommand{\ba}{\begin{align}}
\newcommand{\h}{{(h)}}
\newcommand{\al}{{(\alpha)}}
\newcommand{\ls}{\!\prec\!\!}
\newcommand{\rs}{\!\!\succ\;}
\newcommand{\nol}{:\!}
\newcommand{\nor}{\!:}
\newcommand{\w}{\widetilde}
\newcommand{\der}{\partial}
\title{{\sc Harmonic   measure \& winding of random conformal paths:
 A~Coulomb gas perspective}}
\author{{\sc Bertrand Duplantier}
\\
{\it Institut de Physique Th\'{e}orique, CEA/Saclay}\\
{\it F-91191 Gif-sur-Yvette Cedex, France}\\ \\
{\sc Ilia A. Binder}\\
{\it Department of Mathematics, University of Toronto}\\
{\it Toronto, Ontario, M5S 3G3, Canada}
}

\date{February 14, 2008}

\maketitle

\begin{abstract}

We consider random conformally invariant paths in the complex plane (SLEs). Using the Coulomb
gas method in conformal field theory, we rederive the mixed multifractal exponents
associated with both the harmonic measure and winding (rotation or monodromy) near such critical curves,
previously obtained by quantum gravity methods. The results also extend to the general cases of harmonic
measure moments and winding of multiple paths in a star configuration.

\end{abstract}

\newpage

\tableofcontents

\newpage
\section{Introduction}
\subsection{Historical perspective}
The subject of conformally invariant (CI) random curves in two dimensions has seen spectacular progress in
recent years thanks to the invention of the Stochastic Loewner Evolution (SLE) \cite{schramm1,stflour,lawlerbook}.
This represents the crowning achievement of studies of 2D critical systems undertaken more
than thirty-years ago. The
 first breakthrough came with the introduction of the so-called Coulomb-gas (CG) formalism. The critical properties of
 fundamental two-dimensional statistical models, like the $O(N)$ and Potts models, could then
find an analytic description within that formalism, which led to a profusion of exact results \cite{dennijs,nien}.

This was soon followed by the conformal invariance breakthrough
that occurred in 1984 with the celebrated BPZ article \cite{BPZ}. It was followed by innumerable studies in conformal
field theory
(CFT), which became an essential source of applications to 2D statistical mechanics \cite{WS,cardylebowitz}.
That finally caught the attention of mathematicians \cite{langlands}, first through the peculiar cases of Cardy's formula for
crossing probabilities in percolation \cite{cardy3} and of intersection properties of planar Brownian paths \cite{duplantier2},
resulting several years later in the advent of the SLE era \cite{lawler4,lawler5,smirnov1}.

Meanwhile, the deep relationship between the CG and CFT approaches was brought to light when the Coulomb gas representation
appeared as an explicit
 model for a continuum  of two-dimensional CFTs \cite{feigin,DF}.  A Gaussian free field theory in a given domain,
 modified by a background charge $2\alpha_0$
 (the ``charge at infinity'') that couples the field
 to the domain or boundary curvature, provides a concrete representation of abstract conformal field theories with central charge
 $c=1-24\alpha^2_0\leq 1$ \cite{DF}. Many of the critical properties of statistical models
 could then be obtained from the fusion of these approaches.

 Another breakthrough came in 1988 with the intrusion in statistical mechanics of two-dimensional
 quantum gravity (QG) \cite{KPZ}. The famous KPZ relation between conformal weights in presence of a fluctuating metric and those
  in the Euclidean complex plane \cite{KPZ,david2} could then be checked by explicit
  calculations \cite{kazakov,DK,kostovgaudin}. Though it is not the subject
  of this article, one cannot avoid mentioning in passing the enormous body of knowledge accumulated since then in the related
  field of random matrix theory.

A decade later, just before the advent of SLE, it became clear, first from the reinterpretation of an independent
rigourous study  in probability theory on
intersection properties of planar Brownian paths \cite{lawler2}, that an underlying quantum gravity structure played
an unifying role
in conformal random geometry in the complex plane \cite{duplantier7}. In particular,  revisiting QG allowed the prediction of
 fine geometrical, i.e, multifractal, properties of random critical curves.

 Those concerned the
multifractal spectra associated with the moments of the harmonic measure, i.e., the electrostatic potential,
near CI curves, which could be
derived exactly within that approach for any value of the central charge $c$ \cite{duplantier11}.
A generalization concerned the peculiar {\it mixed} multifractal spectrum  that describes both the
harmonic measure moments and the indefinite winding or rotation (i.e., as logarithmic spirals)  \cite{binder} of a
CI curve about any of its points or of the Green lines of
the potential \cite{DB}.
Higher multifractality spectra
were also introduced and calculated, concerning multiple moments of the harmonic measure and winding in
various sectors of multiple random (SLE) paths in a star configuration \cite{BDjsp, BDMan}.

All these studies resorted to quantum gravity, by
 using a probabilistic representation of harmonic measure moments in terms of collections of Brownian paths,
and by performing a {\it ``transmutation''} of the latter paths into multiple (mutually-avoiding) SLEs, the rules of which are
 established within the QG formalism. The mixed harmonic-rotational spectrum was then obtained by blending this method with
 an earlier Coulomb gas study of the winding angle distributions of critical curves \cite{BDHSwinding}.

 In mathematics, the same multifractal harmonic measure or mixed spectra are the
 subject of present studies and can be obtained rigorously via the probabilistic SLE approach \cite{IABD,Stockholm}.

Recently, the same problem of the geometrical
properties of critical curves was addressed in the physics literature via the Coulomb gas approach alone \cite{BR,RB}.
That work in particular used the statistical equivalence between correlation functions
 of conformal operators
in the complex plane $\mathbb C$ and  correlation functions of a subset of these operators in presence of SLEs,
i.e., in $\mathbb C$ cut by the latter \cite{BB0,cardySLE,NK}. Riemann uniformizing
conformal maps were then
devised to unfold the random paths onto the outside of some smooth domain, e.g., the unit disk.
Standard CFT transformation rules of  primary conformal operators were then the tool of choice.
They allowed in particular to recover the multifractal exponents associated with harmonic measure moments
 near critical curves, originally obtained from quantum gravity.

This ingenious approach thus
raised the
interesting open question of how to  generalize it to mixed  {\it harmonic and winding} properties of
random conformal curves, and
recover the associated mixed multifractal spectra. The aim of the present article is to present such a generalization,
still within the sole Coulomb gas formalism.

Not surprisingly, the main ingredient is the use of arbitrary {\it chiral} primary operators,
 instead of the more familiar
spinless ones. Their associated CFT transformation rules under arbitrary conformal maps
will then yield information about the {\it wild} rotations that occur along Green lines near conformally invariant
random paths.

However, a  technical subtlety will arise here, since one can no longer use only correlation functions
of operators to extract geometrical
information about windings of random paths. To study such windings, one has instead
to resort to {\it products of vertex operators},
without statistical averaging, and study their
  {\it monodromy} properties at short distance. These encode the asymptotic geometrical rotation properties of the
  random paths and
  can be analyzed via operator product expansions, as will be shown below.

\subsection{Harmonic measure \& winding moments}
\subsubsection*{Definition}
Specifically, we  consider a random
conformal path (SLE) $\mathcal S$ and
 the harmonic measure $\omega(0,|z|)$ of a ball of radius $|z|$ centered at point $0\in \mathcal S$, together with the
 (possible indefinite) rotation angle $\vartheta(z)$ of the Green lines (i.e., electrostatic field lines)
 or, equivalently, of the equipotentials, when
 a point $z$
 tends to $0$ while avoiding  $\mathcal S$. We shall evaluate the asymptotic behavior for $z\to 0$ of the
mixed moment,
\begin{eqnarray}
 \omega^{h}(|z|)\,\exp(-p\vartheta(z)) \sim |z|^{\hat x(h,p)},
 \label{xhat}
\end{eqnarray}
when one averages over configurations of the random path. The critical exponent $\hat x(h,p)$ depends on the two
 parameters $h$ and $p$ explicitly \cite{DB}.

 It is also interesting to consider in general multiple SLE paths
 with a star topology, and to define in a similar way the multiple mixed moments \cite{BDMan}:
\begin{eqnarray}
\prod_{i}\omega^{h_i}_i(|z|)\times \exp(-p\vartheta(z)) \sim |z|^{\hat x(\{h_i\};p)},
\end{eqnarray}
where now $\omega_i(|z|)$ is the harmonic measure in each sector $i$ of the star, within a ball of radius $|z|$
centered at apex $0$,
while $\vartheta(z)$ is the common monodromy angle describing
the rotation for $z\to 0$ of the Green lines between the arms of the random conformal star. Here again,
the critical exponent ${\hat x(\{h_i\},p)}$ depends explicitly on all arbitrary exponents $h_i$ and rotation parameter $p$
 \cite{BDMan}, as we shall see in section \ref{S3}.
\subsubsection*{Riemann map}
A known way of computing these moments is to consider the conformal map $w(z)$ which transforms the
exterior of the random path $\mathcal S$ into the exterior of the unit disk $\mathbb D$. The derivative of this map,
$w'(z)$, encodes all the relevant geometrical information. For $z \to 0$,
one has the equivalence $\omega(z)\sim |z w'(z)|$, while the winding angle is given asymptotically
by $\vartheta(z)\simeq -\arg w'(z)$. Thus the mixed moments above can be studied as well
via the moments associated with the derivative of the conformal map
\begin{eqnarray}
 |w'(z)|^{h}\,\exp[p\arg w'(z)] \sim |z|^{x(h,p)},
 \label{xxhat}
\end{eqnarray}
with now an obvious shift of the critical exponent $x(h,p):=\hat x(h,p)-h$.

In the case of a star, one takes a set of points $z_i$ each in  a separate sector, and with
 distances $|z_i|\sim |z|$  all scaling in the same way. Then the windings in each sector are equivalent so
 that  for all $i$, $\arg w'(z_i)\simeq -\vartheta(z)$, whence
\begin{eqnarray}
\label{wom}
\prod_{i}|w'(z_i)|^{h_i}\exp[p_i \arg w'(z_i)]\sim |z|^{x(\{h_i\};p)},
\end{eqnarray}
where  $x(\{h_i\};p):=\hat x(\{h_i\};p)-\sum_ih_i$ and $p=\sum_ip_i$.\\

  To keep the formalism and technical notations to a minimum, and to avoid confusing the reader, we have simply adopted
  the notations of previous work by the Chicago group \cite{RB}, of which the present study can be considered as an extension.
  The reader is thus referred to their article which contains many relevant introductory details. The connection with our previous
  results and notations more familiar in quantum gravity will be recovered at the end of this article.
\section{Derivation}
\subsection{Star \& operator products}
\subsubsection*{Vertex functions}
In complex Gaussian free field theory \cite{Dot,PDF},
the basic objects considered here are ``operator products'', i.e., products of so-called
vertex functions:
\ba  \mathbb P:=\left[\prod_{i=1}^{n}
\mathcal{O}_{\alpha_i',\bar \alpha_i'}(z_i, \bar z_i)
\right]_{\mathbb C},
\label{OpProduct}
\end{align}
where each vertex function or ``operator'' $\mathcal O_{\alpha,\bar \alpha}(z, \bar z)$ is formally
made of two respectively holomorphic and anti-holomorphic components
\ba\mathcal O_{\alpha,\bar \alpha}(z, \bar z)&=  V^\alpha(z) \times {\bar V}^{\bar\alpha}(\bar z)\label{Oaa}\\
 V^\alpha(z)& :=
e^{i\sqrt 2 \alpha \phi(z)}, \qquad {\bar V}^{\bar\alpha}(\bar z) :=
e^{i \sqrt2 \bar\alpha \bar\phi(\bar z)}, \nonumber
\end{align}
with Gaussian free-field correlators:
\begin{eqnarray}
\langle \phi(z)\phi(z')\rangle=-\log(z-z'),\langle \bar \phi(\bar z)\bar \phi(\bar z')\rangle=-\log(\bar z-\bar z'),
\langle \phi(z)\bar \phi(\bar z')\rangle=0.
\la{Gausscorr}
\end{eqnarray}
The holomorphic and the anti-holomorphic weights of the vertex
operators are found in a standard way by applying the stress-energy tensor
to them: \ba h=h_\alpha &= \alpha(\alpha -
2\alpha_0), & {\bar h}={h}_{\bar \alpha} &= \bar\alpha (\bar\alpha -
2\alpha_0),\label{halpha}
\end{align}
where $2\alpha_0$ is the background charge acting in the Coulomb gas representation of the Gaussian free field theory.
A given weight $h$ thus corresponds to two possible charges:
\begin{eqnarray}
\alpha_h=\alpha_0\pm\sqrt{\alpha_0^2+h},
\label{ah}
\end{eqnarray}
an equation which applies separately to holomorphic and anti-holomorphic components.
A vertex operator is spinless (meaning that $h = \bar h$) if either
$\bar\alpha = \alpha$ or $\bar\alpha = 2\alpha_0 - \alpha$. In the sequel, since
we are interested in the rotation of conformally invariant curves, the main tool is the consideration of
  {\it chiral} operators, as opposed to the usual case of spinless ones.
Therefore we explicitly consider in (\ref{OpProduct}) operators $\mathcal{O}_{\alpha_i',\bar \alpha_i'}$
with different conformal
weights:

$$h'_i:=h_{\alpha_i'}\neq h_{\bar \alpha_i'}=:\bar h'_i.$$

\subsubsection*{Star operator product}
To study the harmonic or rotation properties of a star configuration of $n$ critical curves, i.e., SLEs, the
main mathematical object is the {\it star operator product} \cite{RB}
\ba \mathbb P_n(\{z_i, \bar z_i\}) :=  \left[\Psi_{0,n/2}(0)\prod_{i=1}^{n}
\mathcal{O}_{\alpha_i',\bar \alpha_i'}(z_i, \bar z_i)
\right]_{\mathbb C},
\label{StarProduct}
\end{align}
where the operator $\Psi_{0,n/2}(0)$ in (\ref{StarProduct}) is, in the conformal field theory parlance, the
operator corresponding to the existence of $n$ critical curves originating at point $z=0$ in the plane.
It represents the seed of a star configuration
of $n$ such curves, i.e., SLEs. The {\it test operators} $\mathcal{O}_{\alpha_i',\bar \alpha_i'}$ are a priori chiral, so
that $\alpha_i'\neq \bar\alpha_i'$ (or $\neq 2\alpha_0-\bar\alpha_i'$).

We shall need the standard CG notations for holomorphic charges \cite{PDF}:
\begin{eqnarray}\alpha_{r,s} :=
\alpha_0 - \frac{1}{2}(r \alpha_+ + s \alpha_-) =
\frac{1}{2}(1-r)\alpha_+ + \frac{1}{2}(1-s)\alpha_-, \label{ars}
\end{eqnarray}
where the basic charges are given in terms of the SLE parameter $\kappa\in [0,8]$ by \cite{RB}
\ba
\alpha_+ &= \frac{\sqrt\kappa}{2}, \quad \alpha_- = -
\frac{2}{\sqrt\kappa}. \la{alphas}\\
2\alpha_0 &=\alpha_++\alpha_-=\frac{\sqrt\kappa}{2}-\frac{2}{\sqrt\kappa}.\label{a0k}
\end{align}
The charges $\alpha_\pm$ satisfy the simple relations
$$\alpha_\pm =
\alpha_0 \pm \sqrt{\alpha_0^2 + 1},$$
thus both correspond to a conformal weight $h_{\alpha_\pm}=1$, leading to the possibility of using them
as fundamental ``screening charges'' to build up the algebra of screening operators \cite{DF}.

The central charge of the CFT associated with the Gaussian free field theory, modified
{\it \`a la Feigin-Fuchs \& Dotsenko-Fateev} by the
background charge $2\alpha_0$,
is given by \ba c= 1 - 24 \alpha_0^2 = 1 - \frac{3}{2} \frac{(\kappa -
4)^2}{\kappa}, \label{cc}
\end{align}
while the conformal weights corresponding to charges $\alpha_{r,s}$ are 
\begin{eqnarray}
h_{r,s} := \alpha_{r,s}(\alpha_{r,s} - 2\alpha_0) = \frac{(r \kappa
- 4 s)^2 - (\kappa - 4)^2}{16 \kappa}. \la{Kac weights}
\end{eqnarray}

In this representation,
 the holomorphic (or anti-holomorphic) charge of the bulk curve-creating
operator $\Psi_{0,n/2}$ can be taken as
 $\alpha_{0, n/2}=\alpha_0 - \frac{n}{4} \alpha_-$ (with conjugate $\alpha_{0,- n/2}$),
and corresponds in the Coulomb gas formalism to a combination of electric and magnetic charges \cite{RB,DS}.
The corresponding operator is {\it spinless} with conformal
 weight \cite{DS}\ba h_{0,n/2} = \frac{4n^2 - (\kappa - 4)^2}{16 \kappa}.
\label{h0n}
\end{align}
Notice that operator products such as (\ref{OpProduct}) or (\ref{StarProduct}) are understood as objects to be inserted
in correlation functions, in the same way as the formal holomorphic {\it vs} anti-holomorphic factorization (\ref{Oaa})
is meaningful only
in such correlations (\ref{Oaa}). Now,
in the Coulomb gas formulation of conformal field theory, correlation functions are to be
evaluated for a set of charges which overall respect electroneutrality, so that the sum of the latter
always equals $-2\alpha_0$,
to compensate for the existence of the background charge $2\alpha_0$. So ultimately,
the set of charges introduced by the operator product $\mathbb P_n$ near the origin, namely the set
 $\{ \alpha_{0, n/2},\,(\alpha_i',\bar \alpha_i'), i=1,\cdots n\}$
 has to be compensated by other charges located at the observation 
 points of the other
 remaining vertex functions.

 When dealing with ``conformal blocks'', namely
treating separately the holomorphic and anti-holomorphic parts in correlation functions, electroneutrality
should apply separately to both sectors. Furthermore, if holomorphic (or anti-) electroneutrality does
not apply in the arguments of the correlation functions of the original vertex operators, it is known that
 supplementary integral {\it screening operators}, with screening charges $\alpha_+$ and $\alpha_-$ (\ref{alphas}), can be introduced
 to extend the domain of definition of the theory.

 This caveat here is to explain why, when dealing
 with local operator products, such as that (\ref{OpProduct}) or (\ref{StarProduct}), one does not have to
 enforce Coulomb gas
neutrality rules locally, but only globally. We are here actually concerned only
with the short distance and monodromy properties of the star operator product (\ref{StarProduct})
near the origin, which result from the analysis of the short-distance expansion of chiral
operators $\mathcal O_{\alpha',\bar\alpha'}$ in presence of the spinless curve-creating one $\Psi_{0,n/2}$, and
the CG formalism will
be kept at a minimum. It would be interesting to make a more complete study of the algebra of holomorphic
operators, including screening ones, for the present problem and in the context of SLE theory \cite{DF,JSSK}.
\subsection{Star operator product expansion}
\subsubsection*{OPE for vertex functions}
Let us first consider an arbitrary  product of vertex functions in the complex plane:
\ba \mathbb P =  \prod_{i\in \mathcal I}\left[
\mathcal{O}_{\alpha_i',\bar \alpha_i'}(z_i, \bar z_i)\right]_{\mathbb C}
= \prod_{i\in \mathcal I}\left[
V_{\alpha_i'}(z_i)\times \bar V_{\bar \alpha_i'}(\bar z_i)\right]_{\mathbb C},
\label{prod}
\end{align}
where we have made explicit the formal factorization into holomorphic and anti-holomorphic vertex operators.

When a subset $\mathcal P$ of $\mathcal I$
of points $i\in \mathcal P$ is {\it contracted} towards a common point, like the origin here, i.e,
$\forall i\in \mathcal P, z_i\to 0$, it is well-known that the limiting object is given
by the so-called  {\it ``operator product expansion''} (OPE), with coefficients that are singular functions
of the set of points $\mathcal P$.

 In a Gaussian free-field theory, and for a product of vertex functions, i.e., complex exponentials of the field, these
coefficients are precisely given by the average correlators of the set of vertex functions located at the contracted
set $\mathcal P$. The simplest and  most convenient
 way to express this result is probably to use the {\it normal ordering} of operators $:(\cdots):$ such that \cite{BS,DDG}
\begin{eqnarray}
 \prod_{i\in \mathcal P}\left[
\mathcal{O}_{\alpha_i',\bar \alpha_i'}(z_i, \bar z_i)\right]_{\mathbb C}
= \langle\prod_{i\in \mathcal P}\left[\mathcal{O}_{\alpha_i',\bar \alpha_i'}(z_i, \bar z_i)\right]\rangle_{\mathbb C}\,\, :\prod_{i\in \mathcal P}\left[
\mathcal{O}_{\alpha_i',\bar \alpha_i'}(z_i, \bar z_i)\right]_{\mathbb C}:\, .
\label{no}
\end{eqnarray}
In this notation, the vertex operators appearing inside  $:(\cdots):$, when inserted in any correlation function, are
to be Wick-contracted  only with operators located {\it outside} of contracting set $\mathcal P$.

In the complex plane, because of the form (\ref{Gausscorr}) of Gaussian averages
(or ``Wick contractions'', or also ``free-field propagators''), the holomorphic and anti-holomorphic sector contributions 
to a correlation function decouple, hence factorize:
\begin{eqnarray}\nonumber
\langle\prod_{i\in \mathcal P}\mathcal{O}_{\alpha_i',\bar \alpha_i'}(z_i, \bar z_i)\rangle_{\mathbb C}
&=&\langle\prod_{i\in \mathcal P}\left[V_{\alpha_i'}(z_i) \bar V_{\bar \alpha_i'}(\bar z_i)
\right]\rangle_{\mathbb C}\\
&=&\prod_{i,j\in \mathcal P, i<j}
(z_i - z_j)^{2\alpha_i' \alpha_j'}  (\bar z_i - \bar z_j)^{2\bar \alpha_i' \bar \alpha_j'}.
\end{eqnarray}
This in turn gives the explicit form of OPE (\ref{no}).

\subsubsection*{Star OPE}
Applying the above OPE result to the star-operator product
\ba \mathbb P_n(\{z_i, \bar z_i\}) :=  \left[\Psi_{0,n/2}(0)\prod_{i=1}^{n}
\mathcal{O}_{\alpha_i',\bar \alpha_i'}(z_i, \bar z_i)
\right]_{\mathbb C},
\label{starOP}
\end{align}
when the set of points $z_i, i=1, \cdots,n$ all contract to the origin $z=0$, yields
\ba \mathbb P_n(\{z_i, \bar z_i\}) &=
\prod_i z_i^{2\alpha_{0, n/2} \alpha_i'} \bar z_i^{2\alpha_{0, n/2} \bar \alpha_i'}\prod_{i<j} (z_i - z_j)^{2
\alpha_i' \alpha_j'}  (\bar z_i - \bar z_j)^{2
\bar \alpha_i' \bar \alpha_j'}\label{hol}\\\nonumber &
\times :\mathbb P_n(\{z_i, \bar z_i\}):\,,
\end{align}
 a result that will be our main tool in the analysis of the multifractal properties of a set of critical
 paths in a star configuration.

 \subsection{Monodromy \& harmonic measure of a single path}
 \la{single}
 \subsubsection*{Original operator product}
\begin{figure}[htbp]
\centering
\includegraphics[width=0.3290\textwidth]{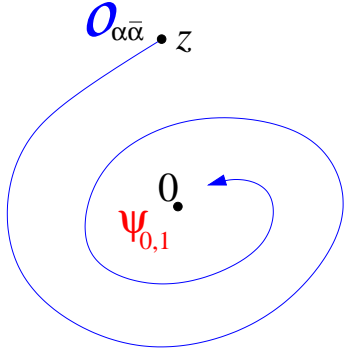}
\caption{Curve-creating operator $\Psi_{0,1}(0)$ and chiral test operator $O_{\alpha,\bar\alpha}(z,\bar z)$, with
monodromy of $z$ around base point $0$.}
\la{fig.spirale}
\end{figure}
Let us consider the standard case, where the star reduces to a single random
curve passing through the origin  $z=0$, with $n=2$ arms, i.e., a set of two semi-infinite curves arising at the origin.
 The associated operator is $\Psi_{0,1}$ with holomorphic charge 
 $\alpha_{0, 1}=\alpha_+/2.$

Let us place first one arbitrary ``test'' operator $\mathcal{O}_{\alpha',\bar \alpha'}(z, \bar z)$ near the origin, so
as to ``measure'' the harmonic measure moments and rotation on one side of the critical curve ({Fig.}\ref{fig.spirale}).
The case of two test operators on both sides of the path
will be treated later as the peculiar $n=2$ case of the general $n$-star geometry.
 In the present geometrical situation,  the relevant
operator product is
\ba  \mathbb P(z,\bar z):= \left[\Psi_{0,1}(0)
\mathcal{O}_{\alpha',\bar \alpha'}(z, \bar z)
\right]_{\mathbb C},
\label{corr1}
\end{align}
and  we are interested in its rotation, i.e., monodromy properties at short-distance (Fig.\ref{fig.spirale}).
Its short-distance behavior when $z\to 0$ is a simple case of (\ref{hol})
\ba
\mathbb P(z,\bar z)=
 z^{2\alpha_{0, 1} \alpha'} {\bar z}^{2\alpha_{0,1} \bar \alpha'} :\mathbb P(z,\bar z):\, ,
\label{hol1}
\end{align}
such that the SDE coefficient (\ref{hol1}) is explicitly
\ba
\mathbb P(z,\bar z) \sim
 z^{ \alpha_+\alpha'} {\bar z}^{ \alpha_+\bar \alpha'}=(z\bar z)^{ \alpha_+(\alpha'+\bar\alpha')/2} \left(\frac{z}{\bar z}
 \right)^{ \alpha_+(\alpha'-\bar \alpha')/2}.
\label{hol1bis}
\end{align}
\subsubsection*{Operator in path geometry}
\label{conformalmap}
\begin{figure}[htbp]
\centering
\includegraphics[width=0.7\textwidth]{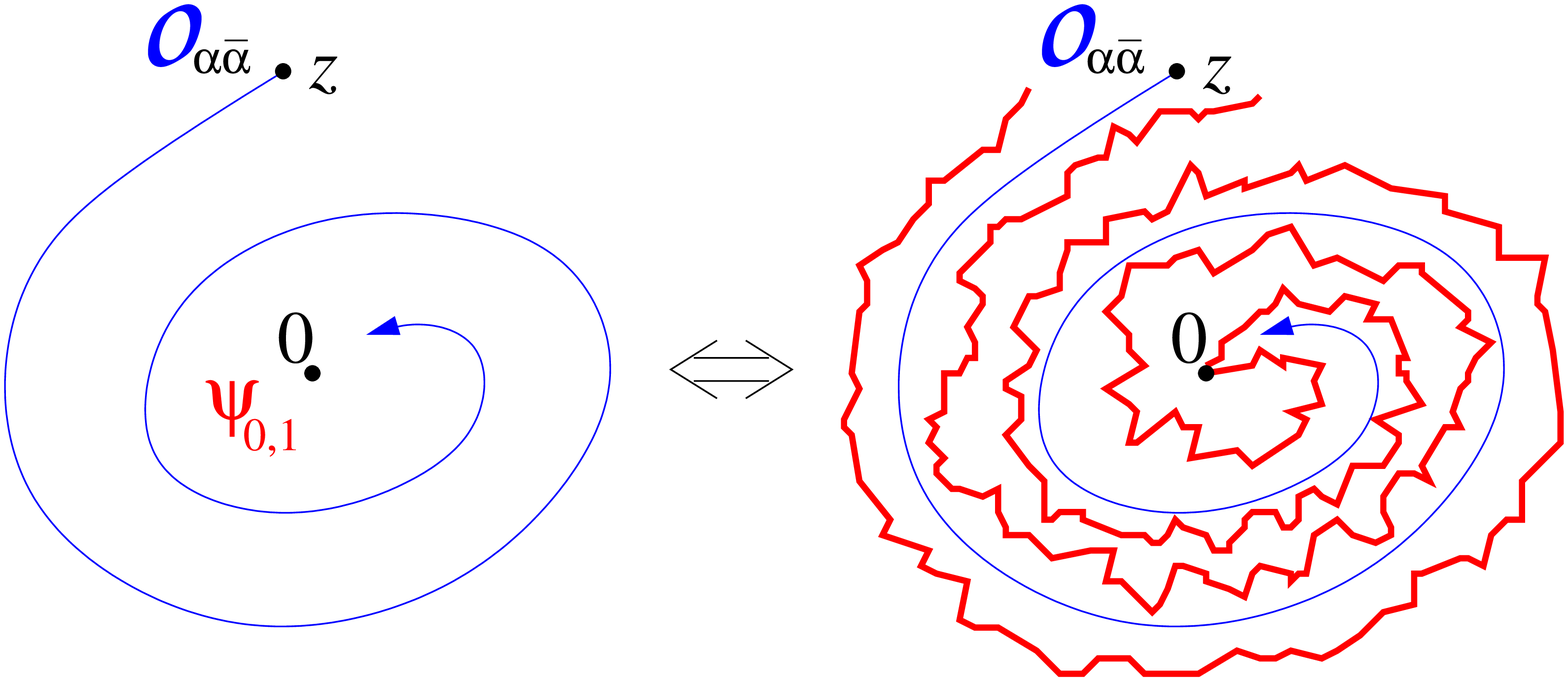}
\caption{Illustration of the statistical identity (\ref{OPE1}) of the product of operators $\Psi_{0,1}$ and
$\mathcal{O}_{\alpha,\bar \alpha}$ with the latter one put in presence of the stochastic path.}
\la{fig.spirale4}
\end{figure}
One first writes the identity {\it in law} of the operator product $\mathbb P$ (\ref{corr1}) in $\mathbb C$,
of the path creating vertex operator $\Psi_{0,1}$
by the test operator, to the same test operator
 but now in presence of the
random path $\mathcal S$ originally represented by $\Psi_{0,1}$,
  hence to the same test operator in the
complex plane ${\mathbb C}  \backslash {\mathcal S}$ slit by the curve:
\ba  \mathbb P(z,\bar z):=\left[\Psi_{0,1}(0)
\mathcal{O}_{\alpha',\bar \alpha'}(z, \bar z)
\right]_{\mathbb C}\stackrel{\rm (in\,\, law)}{=}\left[ \mathcal{O}_{\alpha',\bar \alpha'}(z, \bar z)
\right]_{{\mathbb C}\backslash{\mathcal S}}.
\label{OPE1}
\end{align}
This identity in law means that averaging within correlation functions the left-hand side over the complex Gaussian free field (GFF), or the
right-hand side over the GFF in presence
of $\mathcal S$ and over the configurations of the random path $\mathcal S$ yield the same result \cite{BB0,RB}.
[As discussed in \cite{RB}, the precise boundary conditions on $\mathcal S$ for the geometrical random fields associated with the complex GFF may depend on the phase of the critical system
(``dilute'' for simple SLE paths, $\kappa \leq 4$, or ``dense'' for $\kappa >4$), and are not specified here.
In the holomorphic formalism, analytic continuation in $\kappa$ allows one to pass from one phase to the other.]

\begin{figure}[htbp]
\centering
\includegraphics[width=0.7\textwidth]{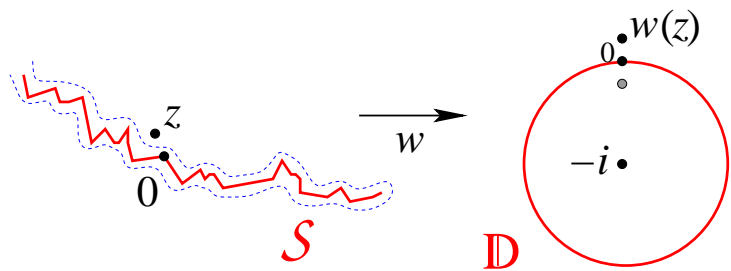}
\caption{Conformal map of the complement $\mathbb C\backslash \mathcal S$
of the random path $\mathcal S$ in $\mathbb C$ to the exterior of the unit disk $\mathbb D$. A mirror image of $w(z)$
by inversion  with respect to the unit circle $\partial\mathbb D$ appears underneath the boundary.}
\la{fig.spirale2}
\end{figure}
\subsubsection*{Conformal map}
The complex plane slit by $\mathcal S$ has the topology of the disk. One introduces the
Riemann uniformizing map $z\to w(z)$ that opens the slit $\mathcal S$ into the unit disk $\mathbb D$
centered at $-i$, so that $w(0)=0$ (Fig.\ref{fig.spirale2}) \cite{BR,RB}. This map
naturally depends on the random path $\mathcal S$. Under this conformal map, vertex operators transform like primary fields,
whence \cite{PDF}: \ba
{\left[\mathcal{O}_{\alpha',\bar \alpha'}(z, \bar z)
\right]_{{\mathbb C}\backslash{\mathcal S}}= \left(w'(z)\right)^{h'}  \left(\overline{ w'(z)}\right)^{\bar h'}
 \left[\mathcal{O}_{\alpha',\bar \alpha'}\big(w(z), \overline{w(z)}\big)
\right]_{{\mathbb C\backslash\mathbb D}}},
\label{C-2bis}
\end{align}
where the holomorphic and anti-holomorphic weights of operator $\mathcal{O}_{\alpha',\bar \alpha'}$ are respectively
\begin{equation}
\label{haa'}
h'=h_{\alpha'}=\alpha'(\alpha'-2\alpha_0),\quad
\bar h'=h_{\bar\alpha'}=\bar\alpha'(\bar \alpha'-2\alpha_0),
\end{equation}
such that $$\alpha'=
\alpha_{h'}:=\alpha_0\pm\sqrt{\alpha_0^2+h'}.$$
Associating together the holomorphic and anti-holomorphic weights $h'$ and $\bar h'$, we can write identically
\ba\left[\mathcal{O}_{\alpha',\bar \alpha'}(z, \bar z)
\right]_{{\mathbb C}\backslash{\mathcal S}}=
 {|w'(z)|^{h'+\bar h'}  \left(\frac{d w}{d \overline{w}}(z)\right)^{(h'-\bar h')/2}\left[
 \mathcal{O}_{\alpha',\bar \alpha'}\big(w(z), \overline{w(z)}\big)
\right]_{{\mathbb C\backslash\mathbb D}}},
\label{P1W}
\end{align}
where $$\frac{d w}{d \overline{w}}(z):=\frac{w'(z)}{\overline{w'(z)}}.$$

The final disk configuration in $\left[
 \mathcal{O}_{\alpha',\bar \alpha'}\big(w(z), \overline{w(z)}\big)
\right]_{{\mathbb C\backslash\mathbb D}}$ is a {\it boundary} configuration \cite{JSSK}, where an
image charge appears at the conformal {image} of point ${ w(z)}$ with respect to the disk boundary, namely its image
by inversion with respect to the unit circle. In the limit $z\to 0$, since $w(0)=0$,  the latter image coincides with
the image with respect to the tangent line, i.e., the complex
conjugate $\overline{ w(z)}$, so the same notation is kept here for simplicity. The vertex operator with anti-holomorphic
charge $\bar\alpha'$ in $\left[
 \mathcal{O}_{\alpha',\bar \alpha'}
\right]_{{\mathbb C\backslash\mathbb D}}$ then becomes an {\it holomorphic} vertex function of charge $\bar\alpha'$ taken at image point $\overline{ w(z)}$, so that the holomorphic and anti-holomorphic sectors get coupled
\cite{PDF}.

\subsubsection*{SDE in the $w$ plane}
\label{secSDE}
When $z\to 0$ in the original domain, $w(z)\to 0$ as well as $\overline{ w(z)}\to 0$, so
the two image points pinch the unit circle at the origin in the $w$ plane. The coefficient in the
short-distance expansion (SDE)
of the right-hand side operator in Eq. (\ref{P1W}) is given by the Gaussian
averaged correlation: $$ \big\langle
 \mathcal{O}_{\alpha',\bar \alpha'}\big(w(z), \overline{w(z)}\big)
\big\rangle_{{\mathbb C\backslash\mathbb D}}
=(w(z)\!-\!\overline{w(z)})^{2\alpha'{\bar \alpha'}}. \la{expandedbis}
$$
Then the short-distance expansion of eq.
(\ref{C-2bis}) is \ba \left[ \mathcal{O}_{\alpha',\bar \alpha'}(z, \bar z)
\right]_{{\mathbb C}\backslash{\mathcal S}} \sim\, { |w'(z)|^{h'+\bar h'}
\left(\frac{d w}{d \overline{w}}(z)\right)^{(h'-\bar h')/2}
(w(z)-\overline{w(z)})^{2\alpha'{\bar \alpha'}}}\,.\la{SDE}
\end{align}
After uniformization to the $w$ plane, since the two image points $w(z)$ and $\overline{w(z)}$ pinch the unit circle while staying either in the exterior or the
interior of the disk, they cannot wind about the origin $w(0)=0$ indefinitely for $|z|\to 0$, so $\arg w(z)$ remains {\it bounded},
e.g.,
\begin{equation}
\arg w(z)\in [-\pi, +\pi].
\la{bounded}
\end{equation}
 We conclude that
$w(z)-\overline{w(z)} \sim |z| |w'(z)|$, up to a {\it non-winding} (or non-monodromic) complex phase factor. This finally gives the SDE (\ref{SDE})
\ba \left[\mathcal{O}_{\alpha',\bar \alpha'}(z, \bar z)
\right]_{{\mathbb C}\backslash{\mathcal S}} \sim |z|^{2 \alpha'{\bar \alpha'}} {
|w'(z)|^{h}}
\left(\frac{d w}{d \overline{w}}(z)\right)^{(h'-\bar h')/2}\,, \la{scalingbisbis}
\end{align}
where the overall (harmonic measure) derivative exponent $h$ is defined as
\ba h := h' +\bar h' +
2\alpha'{\bar \alpha'},\la{weightsbis}
\end{align}
Owing to (\ref{haa'}), it is also
\begin{equation}
\label{h}
h=(\alpha'+\bar \alpha')(\alpha'+\bar \alpha'-2\alpha_0).
\end{equation}
It is the weight of an operator whose conformal charge
$\alpha_{h}=\alpha_0\pm\sqrt{\alpha_0^2+h}$ is just $\alpha'+\bar \alpha'$ (or its conjugate):
\begin{equation}
\label{aa'h}
\alpha'+\bar \alpha'=\alpha_{h}=\alpha_0\pm\sqrt{\alpha_0^2+h}.
\end{equation}
Notice also that
\begin{equation}
\label{h-h'}
h'-\bar h'=(\alpha'-\bar \alpha')\left(\alpha'+\bar \alpha'-2\alpha_0\right).
\end{equation}

\subsubsection*{Winding \& uniformizing map}
In the $z$ plane, the curve $\mathcal S$ can wind or rotate indefinitely about any of its points, e.g. about the origin
$0$ when $z\to 0$, so $\arg z\to \pm \infty$. We are  thus
 especially interested in the monodromy properties of the operator $O_{\alpha',\bar{\alpha'}}(z,\bar z)$.

 By analyticity of the conformal map $w$ onto the unit disk, $w(z)\simeq w'(z) z$, so that $\arg w(z)
 \simeq \arg w'(z)+\arg z$. We just have seen in (\ref{bounded}) that $\arg w(z)$ remains bounded, so that asymptotically
 $\arg z \sim -\arg w'(z)$ under the conformal map.

 The winding angle  $\vartheta(z)$ of the Green
 lines of the random curve, asymptotically close to $z=0$
 on the curve, is given by
\begin{equation}
\label{winding0}
\vartheta(z)=\arg z = -\arg w'(z)+\mathcal O(1).
\end{equation}
Therefore one can also write asymptotically 
\begin{equation}
\label{winding}
\vartheta(z)=-\Im \log w'(z)=-\frac{1}{2i}[\log w'(z)-\log {\overline  {w'(z)}}]=-\frac{1}{2i}\log\frac{w'(z)}{{\overline {w'(z)}}},
\end{equation}
so that the exponential winding is
\begin{equation}
\label{expwinding}
\left(\frac{z}{\bar z}\right)^{1/2}=e^{i\vartheta(z)}
\asymp\left(\frac{w'(z)}{{\overline {w'(z)}}}\right)^{-1/2}= \left(\frac{d w}{d \overline{w}}(z)\right)^{-1/2},
\end{equation}
where $\asymp$ here means equality within a {non-winding (non-monodromic)} phase factor. We therefore arrive at the expression for SDE (\ref{scalingbisbis})
\ba \left[ \mathcal{O}_{\alpha',\bar \alpha'}(z, \bar z)
\right]_{{\mathbb C}\backslash{\mathcal S}}  \sim |z|^{2 \alpha'{\bar \alpha'}} {
|w'(z)|^{h}}
e^{-i(h'-\bar h')\vartheta(z)}\,. \la{scalingquater}
\end{align}
\subsubsection*{Mixed moments}
Let us now return to the original OPE (\ref{corr1}), (\ref{hol1}).
 We can rewrite  (\ref{hol1}), (\ref{hol1bis}) as a complex scaling
\ba
 \mathbb P(z,\bar z)=\left[\Psi_{0,1}(0)
\mathcal{O}_{\alpha',\bar \alpha'}(z, \bar z)
\right]_{\mathbb C} \sim
 |z|^{ \alpha_+(\alpha'+\bar\alpha')} e^{i\alpha_+(\alpha'-\bar \alpha')\arg z}.
\label{hol1quater}
\end{align}
Because of the identity in law (\ref{OPE1}), identifying (\ref{hol1quater})
and the short-distance expansion (\ref{scalingquater}) in the transformed slit domain
yields the equivalence
\begin{equation}
\label{final}
|w'(z)|^{h}
e^{-i [h'-\bar h'+\alpha_+(\alpha'-\bar \alpha')]\vartheta(z)}
\sim |z|^{\alpha_+(\alpha'+\bar\alpha')-2 \alpha'{\bar \alpha'}}\,.
\end{equation}
This scaling equivalence is an SDE result, which can be interpreted as describing  typical statistical behavior,
 also expected to hold true in a weaker form after averaging over the configurations of the stochastic path $\mathcal S$:
\begin{equation}
\label{finalb}
\ls {
|w'(z)|^{h}}
e^{-i s\vartheta(z)}\rs\sim |z|^{x}\, ,
\end{equation}
where
\begin{eqnarray}
\label{p}
s&:=&h'-\bar h'+\alpha_+(\alpha'-\bar \alpha')\\
\label{q}
x&:=&\alpha_+(\alpha'+\bar\alpha')-2 \alpha'{\bar \alpha'}.
\end{eqnarray}
Let us now express exponent $x:=x(h,is)$ solely in terms of weight $h$ and winding conjugate parameter $s$.
Using (\ref{h}) (\ref{aa'h}) and (\ref{h-h'}) we have
 \begin{eqnarray}
\label{p'}
s&=&(\alpha'-\bar \alpha')(\alpha_++\alpha'+\bar \alpha'-2\alpha_0)=(\alpha'-\bar \alpha')(\alpha_+
+\alpha_h-2\alpha_0)\\
\label{q'}
x&=&\alpha_+\alpha_h-2 \alpha'{\bar \alpha'}.
\end{eqnarray}
We can further write trivially
$$
4\alpha'{\bar \alpha'}=(\alpha'+\bar\alpha')^2-(\alpha'-\bar\alpha')^2=\alpha_h^2-(\alpha'-\bar\alpha')^2,
$$
to eliminate $\alpha'-\bar\alpha'$ between (\ref{p'}) and (\ref{q'})
 \begin{eqnarray}
\label{pq}
\alpha'-\bar \alpha'&=&\frac{p}{\alpha_+
+\alpha_h-2\alpha_0}\\
\label{p'q'}
x(h,is)&=&\alpha_+\alpha_h-\frac{1}{2}\alpha_h^2+\frac{1}{2}\frac{s^2}{(\alpha_+
+\alpha_h-2\alpha_0)^2}.
\end{eqnarray}
To get the {\bf mixed harmonic measure-rotation exponents}  one has first to {\it analytically continue
$s$:} $s = -ip$ in (\ref{finalb}) so that
\begin{equation}
\label{finalbis}
 \ls {
|w'(z)|^{h}}
e^{- p\vartheta(z)}\rs\,\,\sim |z|^{x(h,p)},
\end{equation}
with now
\begin{eqnarray}
\label{q''}
x(h,p)=\alpha_+\alpha_h-\frac{1}{2}\alpha_h^2-\frac{1}{2}\frac{p^2}{(\alpha_+
+\alpha_h-2\alpha_0)^2}.
\end{eqnarray}
Finally, one has to choose the value
of $\alpha_h$ that vanishes with $h$, namely the ``dilute phase'' or simple SLE path one,
where $\kappa\leq 4$ and $\alpha_0\leq 0$ (see (\ref{a0k}))
\begin{eqnarray}
\label{ahdil}
\alpha_h=\alpha_0+\sqrt{\alpha_0^2+h}.
\end{eqnarray}
The exponent $x(h,p)$ above is identical to the scaling exponent obtained
by quantum gravity in \cite{DB}
for the $h$th power $\omega^h(z)\,e^{-p\vartheta(z)}$ of the harmonic measure $\omega$ with
rotation conjugate parameter $p$
 (up to a natural shift by $h$, due to the local scaling
$\omega(z) \sim |z| |w'(z)|$ for the harmonic measure $\omega(z)$ in a ball of radius $|z|$, cf.
Eqs. (\ref{xhat}) (\ref{xxhat})).
[See section \ref{S3}.]

\subsection{Monodromy \& harmonic measure of multiple paths}
\subsubsection*{Operator product in star geometry}
We first use the identity {\it in law} of the star-operator product (\ref{starOP}) with the operator product
of the test vertex operators in presence of the stochastic star $\mathcal S_n$ in the complex plane:
\begin{eqnarray}
\label{inlaw}
\mathbb P_n(\{z_i,\bar z_i\})\stackrel{(\rm in\,law)}{=}\left[\prod_i \mathcal{O}_{\alpha_i',\bar \alpha_i'}(z_i, \overline{z_i})\right]
_{{\mathbb C\backslash \mathcal  S_n}}.
\end{eqnarray}
\begin{figure}[htbp]
\centering
\includegraphics[width=0.7\textwidth]{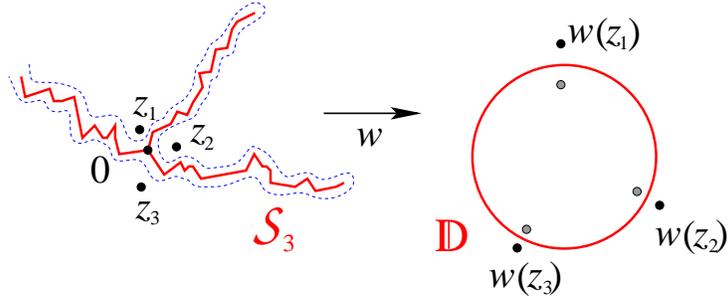}
\caption{Conformal map of the complement $\mathbb C\backslash \mathcal S_3$
of the random 3-star $\mathcal S_3$ in $\mathbb C$ to the exterior of the unit disk $\mathbb D$. Three mirror images
of points $w(z_i), i=1,2,3,$ appear
by inversion  with respect to the unit circle $\partial\mathbb D$.}
\la{fig.spirale3}
\end{figure}
\subsubsection*{Conformal map}
The plane slit by the star having the topology of the disk, a conformal map $w(z)$ transforms the open set
$\mathbb C\backslash\mathcal S_n$ into the exterior of the unit disk $\mathbb D$ \cite{RB}.  The vertex operator product $ {\mathbb P_n}$
(\ref{inlaw}) in presence of $\mathcal S_n$ is transformed, according to primary operator rules,
as:
\begin{eqnarray}\nonumber
\mathbb P_n &=&\left[\prod_i \mathcal{O}_{\alpha_i',\bar \alpha_i'}(z_i, \overline{z_i})\right]
_{{\mathbb C\backslash \mathcal  S_n}}\\
&=& {\prod_i \left(w'(z_i)\right)^{h_i'}  \left(\overline{ w'(z_i)}\right)^{\bar h_i'}
\left[\prod_i \mathcal{O}_{\alpha_i',\bar \alpha_i'}\big(w(z_i), \overline{w(z_i)}\big)\right]
_{{\mathbb C\backslash \mathbb D}}}.\label{OPw}
\end{eqnarray}
Associating for each $i$ the holomorphic and anti-holomorphic weights $h_i'$ and $\bar h_i'$, we can write it identically
\ba
\mathbb P_n = \,\,  {\prod_i |w'(z_i)|^{h_i'+\bar h_i'}  \left(\frac{d w}{d \overline{w}}(z_i)\right)^{(h_i'-\bar h_i')/2}
\left[\prod_i \mathcal{O}_{\alpha_i',\bar \alpha_i'}\big(w(z_i), \overline{w(z_i)}\big)
\right]
_{{\mathbb C\backslash \mathbb D}}}.
\label{P-2}
\end{align}
\subsubsection*{Short-distance expansion}
Now let all points converge to the origin in the original $z$ plane, $z_i\to 0, \forall i=1,\cdots, n$.
As in section (\ref{single}) above, in the disk geometry, the operators on the right hand side of (\ref{P-2}) are
 {\it boundary} operators, each of them made of a pair of holomorphic vertex operators taken at $w(z_i)$ and at its inverted
 mirror image with respect with the unit circle (fig. \ref{fig.spirale3}).  When $z_i\to 0$, each image  coincides
 with the local complex conjugate $\overline{w(z_i)}$ with respect to the local tangent to the unit circle, so the
 same notation is kept by a small abuse of notation.

 As usual, the original anti-holomorphic operators  with  charges $\bar \alpha_i'$ now become {holomorphic} vertex operators 
  located at points $\overline{w(z_i)}$, and get coupled to
 the original holomorphic vertex operators with charges $\alpha_i'$
  located at points ${w(z_i)}$ \cite{PDF}.
 Under the contraction $z_i\to 0, \forall i=1,\cdots,n$,
the SDE of the transformed operator product on the right hand  side of (\ref{P-2}) therefore
 scales as the Gaussian free-field average taken at all pairs of points and images \cite{RB}:
 \ba
&\langle\prod_i \mathcal{O}_{\alpha_i',\bar \alpha_i'}\big(w(z_i), \overline{w(z_i)}\big)
\rangle
_{{\mathbb C\backslash \mathbb D}}=\prod_i
(w(z_i)\!-\!\overline{w(z_i)})^{2\alpha_i'{\bar \alpha_i'}}\la{expanded} \\& \times \prod_{i<j} (w(z_i)
\!-\! w(z_j))^{2 \alpha_i' \alpha_j'} (\overline{w(z_i)}
\!-\! \overline{w(z_j)})^{2 \bar \alpha_i' \bar \alpha_j'}
\prod_{i \neq j} \big(w(z_i)
\!-\! \overline{w(z_j)}\big)^{2 \alpha_i'
\bar \alpha_j'}.\label{Gavallimages}
\end{align}
Among all these factors, the diagonal $i$ terms give the
short-distance behavior. The $i\neq j$ terms are finite in the disk geometry, or give subleading contributions
 for configurations of the star  in which a pair of points $(z_i, z_j)$ lie in
the same sector, since the relative affixes  $w(z_i) - w(z_j)$, $\overline{w(z_i)} - \overline{w(z_j)}$,
$w(z_i) - \overline{w(z_j)}$ then
tend to zero. The short-distance behavior of the operator product
(\ref{P-2}) is thus
\ba \mathbb P_n \sim\, {\prod_i |w'(z_i)|^{h_i'+\bar h_i'}\left(\frac{d w}{d \overline{w}}(z_i)\right)^{(h_i'-\bar h_i')/2}
(w(z_i)-\overline{w(z_i)})^{2\alpha_i'{\bar \alpha_i'}}}.\nonumber
\end{align}
When $z_i\to 0$, the transformed complex coordinate $w(z_i)$ and its image $\overline{w(z_i)}$ pinch the unit circle,
so that $w(z_i)\to\overline{w(z_i)}$ while the winding angle $\arg w(z_i)$ stays {\it bounded}; whence we can simply take:
$w(z_i)-\overline{w(z_i)} \sim |z_i| |w'(z_i)|$, up to a non-monodromic complex phase factor.
This gives the final short-distance behavior of (\ref{P-2})
\ba \mathbb P_n \sim  {\prod_i |z_i|^{2 \alpha_i'{\bar \alpha_i'}}
|w'(z_i)|^{h_i' +\bar h_i' +2\alpha_i'{\bar \alpha_i'}}}
\left(\frac{d w}{d \overline{w}}(z_i)\right)^{(h_i'-\bar h_i')/2}. \la{scaling}
\end{align}
Hence the following exponents  appear for the harmonic measure derivative terms : \ba h_i := h_i' +\bar h_i' +
2\alpha_i'{\bar \alpha_i'}=(\alpha'_i+\bar \alpha_i')(\alpha'_i+\bar \alpha_i'-2\alpha_0),\la{weights}
\end{align}which are the conformal weights of operators whose conformal charges simply result from the addition of
holomorphic and anti-holomorphic charges~:
$$\alpha_{h_i}=\alpha_0+\sqrt{\alpha_0^2+h_i}=\alpha'_i+\bar \alpha_i'.$$

\subsubsection*{Windings \& uniformizing map}
Let us now return to the SDE (\ref{hol}) in the original $z$ plane
\begin{eqnarray} \mathbb P_n (\{z_i,\bar z_i\})\sim
\prod_i z_i^{2\alpha_{0, n/2} \alpha_i'} \bar z_i^{2\alpha_{0, n/2} \bar \alpha_i'}\prod_{i<j} (z_i - z_j)^{2
\alpha_i' \alpha_j'}  (\bar z_i - \bar z_j)^{2
\bar \alpha_i' \bar \alpha_j'}
\label{holn}
\end{eqnarray}
and, as in Eq. (\ref{hol1bis}), let us separate the complex modulus and argument parts, by rewriting it as~:
\begin{eqnarray} \nonumber\mathbb P_n(\{z_i,\bar z_i\})
&\sim&
\prod_i \vert z_i\vert^{2\alpha_{0, n/2} (\alpha_i'+\bar \alpha_i')} \left(\frac{z_i}
{\bar z_i}\right)^{\alpha_{0, n/2}
(\alpha_i'-\bar \alpha_i')}\\&\times&\prod_{i<j} \vert z_i - z_j \vert^{2
\alpha_i' \alpha_j'+2\bar \alpha_i' \bar \alpha_j'}  \left(\frac{z_i - z_j}{\bar z_i - \bar z_j}\right)^{
\alpha_i' \alpha_j'-\bar \alpha_i' \bar \alpha_j'}.
\label{holnsep}
\end{eqnarray}

We are interested in the monodromy properties of expression (\ref{holnsep}) when all points in their own
 sectors converge to the star apex $z=0$ while
avoiding touching the star, which can wind indefinitely about its origin.
Let us then introduce a common scaling $\vert z\vert$ for all distances $\vert z_i\vert$ to the star apex,
as well as a common asymptotic
winding angle $\vartheta(z)$ for all arguments $\arg z_i$:
\begin{equation}
\label{windingi}
z_i=\vert z\vert e^{i\vartheta (z)} \zeta_i,\,\,\,\arg z_i=\vartheta(z) +\arg \zeta_i,
\end{equation}
where all moduli $\vert \zeta_i\vert$ and arguments $\arg \zeta_i$ remain bounded when $|z|\to 0$.
Since trivially
\begin{equation}\label{thetaibis}
\left({z_i}/{\bar z_i}\right)^{1/2}
= e^{ i\vartheta(z)}\left({\zeta_i}/{\bar \zeta_i}\right)^{1/2},
\end{equation}
the short-distance expansion (\ref{holnsep}) obeys the simple identity
\begin{eqnarray} \nonumber
\mathbb P_n(\{z_i,\bar z_i\})
&= &  \vert z\vert^{2\alpha_{0, n/2} \sum_i(\alpha_i'+\bar \alpha_i')+\sum_{i<j}(
2\alpha_i' \alpha_j'+2\bar \alpha_i' \bar \alpha_j')} \\ \nonumber&\times& e^{i\vartheta(z) [2 \alpha_{0, n/2}
\sum_i(\alpha_i'-\bar \alpha_i')+2 \sum_{i<j}
(\alpha_i' \alpha_j'-\bar \alpha_i' \bar \alpha_j')]}\\
&\times&\mathbb P_n(\{\zeta_i,\bar \zeta_i\}).\label{holnsepzeta}
\end{eqnarray}
Let us now consider the geometrical setting in the transformed $w$-plane.
Since all arguments $\arg w(z_i)$ stay bounded there, we have for $z_i\to 0$, as in Eq. (\ref{winding0}),
$\arg w'(z_i)= -\arg z_i+\mathcal O(1)$, so that, as in (\ref{expwinding})
\begin{equation}\label{thetai}
\left(\frac{z_i}{\bar z_i}\right)^{1/2}
\asymp\left(\frac{d w}{d \overline{w}}(z_i)\right)^{-1/2}.
\end{equation}
Identity (\ref{thetaibis}) then yields immediately the common asymptotic behavior
for all $i$
\begin{equation}\label{commonasymp}
\forall i, \,\,\,\left(\frac{d w}{d \overline{w}}(z_i)\right)^{-1/2}\asymp e^{i\vartheta(z)},
\end{equation}
in terms of the unique rotation angle $\vartheta(z)$ (\ref{windingi}) , and up to non-monodromic phase factors.
Applying this to the SDE result (\ref{scaling}) gives the asymptotic formula
\begin{eqnarray}  \mathbb P_n \sim  |z|^{2 \sum_i\alpha_i'{\bar \alpha_i'}} e^{-i\vartheta(z)\sum_i(h_i'-\bar h_i')}
{\prod_i
|w'(z_i)|^{h_i}}.\la{scalingztheta}
\end{eqnarray}
\subsubsection*{Multiple moments and rotation}
Because of the identity in law (\ref{inlaw}), we can now identify expressions (\ref{holnsepzeta}) and
(\ref{scalingztheta}) and get the fundamental scaling formula
for the multiple harmonic measure factors and (indefinitely) rotating phase factor
\begin{eqnarray}\nonumber
 &&
e^{-i\vartheta(z) [\sum_i(h_i'-\bar h_i')+ 2 \alpha_{0, n/2}
\sum_i(\alpha_i'-\bar \alpha_i')+2 \sum_{i<j}
(\alpha_i' \alpha_j'-\bar \alpha_i' \bar \alpha_j')]}\times{\prod_i|w'(z_i)|^{h_i}}\\ \label{fund}
&&\sim  \vert z\vert^{2\alpha_{0, n/2} \sum_i(\alpha_i'+\bar \alpha_i')+\sum_{i<j}(
2\alpha_i' \alpha_j'+2\bar \alpha_i' \bar \alpha_j')-\sum_i 2\alpha_i'{\bar \alpha_i'}}.
\end{eqnarray}
We obtained this scaling relation from a short-distance operator product expansion, which
yields the leading (typical) scaling behaviour of the product of harmonic measure factors and rotation phase factor.  It
can naturally be expected to also give the weaker result on the statistical average $\ls\cdots \rs$ of the same quantity over the configurations of
the multiple random paths of star
${\mathcal S_n}$~:
\begin{eqnarray}\label{eqfund}
 \ls {\prod_i|w'(z_i)|^{h_i}}
e^{-i\vartheta s}\rs
\sim  \vert z\vert^{x(\{h_i\};is)} ,
\end{eqnarray}
with the parametric representation in terms of holomorphic and anti-holomorphic charges
\begin{eqnarray}
h_i&:=&h_i' +\bar h_i' +2\alpha_i'{\bar \alpha_i'}\\
h'_i&=&\alpha_i'(\alpha_i'-2\alpha_0);\,\,\,\bar h'_i=\bar \alpha_i'(\bar \alpha_i'-2\alpha_0)\\
\label{s}
s&:=&\sum_i(h_i'-\bar h_i')+ 2 \alpha_{0, n/2}
\sum_i(\alpha_i'-\bar \alpha_i')+2 \sum_{i<j}
(\alpha_i' \alpha_j'-\bar \alpha_i' \bar \alpha_j')\\
\label{x}
x&:=&2\alpha_{0, n/2} \sum_i(\alpha_i'+\bar \alpha_i')+\sum_{i<j}(
2\alpha_i' \alpha_j'+2\bar \alpha_i' \bar \alpha_j')-\sum_i 2\alpha_i'{\bar \alpha_i'}.
\end{eqnarray}
It remains to find the explicit expression of the scaling exponent $x:=x(\{h_i\};is)$ in terms of the set of
weights $h_i$ and of Fourier variable $s$ conjugate to rotation $\vartheta$. Let us introduce the notations
\begin{eqnarray}\nonumber
h_{\alpha'}&=&\alpha'(\alpha'-2\alpha_0),\,\,\,h_{\bar\alpha'}=\bar\alpha'(\bar\alpha'-2\alpha_0)\\\nonumber
{\alpha'}&:=&\sum_i\alpha_i',\,\,\,{\bar \alpha'}:=\sum_i\bar \alpha_i'\\\nonumber
\alpha_{h_i}&=&\alpha_i'+\bar \alpha_i'\\\nonumber
\alpha_{\{h\}}&:=&\alpha'+\bar \alpha'=\sum_i(\alpha_i'+\bar \alpha_i')=\sum_i\alpha_{h_i}.
\end{eqnarray}
A little bit of algebra first shows that $s$ (\ref{s}) can be written in the compact form
\begin{eqnarray}\nonumber
s&=&h_{\sum_i\alpha_i'}+2\alpha_{0, n/2}\sum_i\alpha_i'-h_{\sum_i\bar\alpha_i'}-2\alpha_{0, n/2}\sum_i\bar\alpha_i'\\
\nonumber
&=&h_{\alpha'}-h_{\bar\alpha'}+2\alpha_{0, n/2}(\alpha'-\bar\alpha')\\
\nonumber
&=&(\alpha'-\bar\alpha')(\alpha'+\bar\alpha'-2\alpha_0+2\alpha_{0, n/2})\\
&=&(\alpha'-\bar\alpha')(\alpha_{\{h\}}-2\alpha_0+2\alpha_{0, n/2}).
\label{s1}
\end{eqnarray}
One can also check that the exponent $x$ (\ref{x}) can be written in the compact form
\begin{eqnarray}\nonumber
x&=&2\alpha_{0, n/2}\sum_i\alpha_{h_i}+\left(\sum_i\alpha_{h_i}\right)^2-\sum_i\alpha_{h_i}^2-2\alpha'\bar\alpha'\\
&=&2\alpha_{0, n/2}\alpha_{\{h\}}+\alpha_{\{h\}}^2-\sum_i\alpha_{h_i}^2-2\alpha'\bar\alpha'.
\label{x1}
\end{eqnarray}
Note that after this compaction of formulae, the expressions (\ref{s1}) and (\ref{x1}) are  similar to expressions
(\ref{p'}) and (\ref{q'}). We therefore eliminate $\alpha'-\bar\alpha'$ in the same way as above and arrive at
a formula similar to (\ref{pq}) and (\ref{p'q'})
\begin{eqnarray}
\label{sx}
\alpha'-\bar \alpha'&=&\frac{s}{\alpha_{\{h\}}-2\alpha_0+2\alpha_{0, n/2}}\\
\nonumber
x(\{h_i\}; is)&=&2\alpha_{0, n/2}\,\alpha_{\{h\}}+\frac{1}{2}\alpha_{\{h\}}^2-\sum_i\alpha_{h_i}^2\\
&+&\frac{1}{2}\frac{s^2}
{(\alpha_{\{h\}}-2\alpha_0+2\alpha_{0, n/2})^2},
\label{s'x'}
\end{eqnarray}
where we recall that $\alpha_{\{h\}}=\sum_i\alpha_{h_i},\,\,\alpha_{h_i}=\alpha_0+\sqrt{\alpha_0+h_i}$,
hence exponent $x$ has now an explicit form in terms of the set of weights $\{h_i\}$ and of $s$.

It remains
to  analytically continue $s$ into $s=-ip$ to get the expectation of the multiple harmonic measure moments and
Laplace transform of the rotation:
\begin{eqnarray}\label{fundanalytic}
 \ls {\prod_i|w'(z_i)|^{h_i}}
e^{-p\vartheta }\rs
&\sim&  \vert z\vert^{x(\{h_i\};p)} ,\\\nonumber
x(\{h_i\}; p)&=&2\alpha_{0, n/2}\,\alpha_{\{h\}}+\frac{1}{2}\alpha_{\{h\}}^2-\sum_i\alpha_{h_i}^2\\
&-&\frac{1}{2}\frac{p^2}
{(\alpha_{\{h\}}-2\alpha_0+2\alpha_{0, n/2})^2}.
\label{xp}
\end{eqnarray}
CG formula (\ref{xp}) for $p=0$  coincides naturally with the result found in ref. \cite{RB}.
\section{Comparison to quantum gravity results}
\label{S3}
In previous work  \cite{BDMan} we introduced for $\rm{SLE}_\kappa$ the KPZ relation
$$h=\mathcal U_\kappa(\Delta):=\frac{1}{4}\Delta\left(\kappa\Delta+4-\kappa\right)$$
 between conformal
weights $\Delta$ in a fluctuating metric with a conformal factor given by a Gaussian free field
(two-dimensional ``quantum gravity'' (QG)) and conformal weights $h$ in the complex plane.
Its inverse reads
$$
\Delta=\mathcal U^{-1}_\kappa(h)=\frac{1}{2\kappa}\sqrt{16\kappa h +(\kappa-4)^2}+
\frac{1}{2}\left(1-\frac{4}{\kappa}\right).
$$
Using (\ref{a0k}) and (\ref{ahdil}) we identify
\begin{eqnarray}\la{aU}
\alpha_h=\frac{\sqrt\kappa}{2}\,\mathcal U^{-1}_\kappa(h).
\end{eqnarray}
The CG results above can be written as
\begin{eqnarray}\nonumber
x(\{h_i\}; p)&=&x(\{h_i\}; 0)
-\frac{\kappa}{2} \frac{p^2}{L^2_{\{h\}}}\\ \nonumber
L_{\{h\}}&:=&\sqrt\kappa\,{(\alpha_{\{h\}}-2\alpha_0+2\alpha_{0, n/2})}\\
x(\{h_i\}; 0)&=&2\alpha_{0, n/2}\,\alpha_{\{h\}}+\frac{1}{2}\alpha_{\{h\}}^2-\sum_i\alpha_{h_i}^2\, .
\label{xp0}
\end{eqnarray}
Using (\ref{aU}) we find their equivalent quantum gravity expressions \cite{BDMan}
\begin{eqnarray}\label{QG}
x(\{h_i\}; p)&=&x(\{h_i\}; 0)
-\frac{\kappa}{2} \frac{p^2}{L^2_{\{h\}}}\\\label{QGL}
L_{\{h\}}&=&\frac{\kappa}{2}\sum_i\mathcal U^{-1}_\kappa(h_i)+n,\\\label{QQG}
x(\{h_i\}; 0)+\sum_i h_i&=&2\,\mathcal U_\kappa\left[\frac{1}{2}\left(\frac{2}{\kappa}L_{\{h\}}+1-
\frac{4}{\kappa}\right)\right]\\
&-&2\,\mathcal U_\kappa\left[\frac{1}{2}\left(\frac{2}{\kappa}n+1-
\frac{4}{\kappa}\right)\right].
\label{QQG0}
\end{eqnarray}
[As seen in (\ref{wom}), the term $\sum_ih_i$ simply comes from the passage from derivative moments to harmonic
measure ones.]

Result (\ref{q''}) for the multifractal exponent $x(h,p)$ of a single-sided SLE simply corresponds
to $n=2$ and $(h_1,h_2)=(h,0)$ in QG formulae (\ref{QG})--(\ref{QQG0}).\\

\noindent The interpretation of these formulae is quite clear:

\noindent $\bullet$ (\ref{QG}) is the Coulomb gas formula obtained
a long time ago for the exponent governing the winding angle distribution
of a star made of a number $L_{\{h\}}$ of random paths (SLEs) \cite{BDHSwinding,DB,BDMan};

\noindent $\bullet$ $L_{\{h\}}$ (\ref{QGL}) represents, in a star topology,
the {\it effective number} of SLEs that are exactly equivalent through QG to a collection of $h_i$, $i=1,\cdots,n,$ Brownian paths
(which represent the set of powers $h_i$ of the harmonic measure), in addition to the $n$ original SLEs \cite{BDjsp,BDMan};

\noindent $\bullet$ The scaling dimension
(\ref{QQG}) is the result obtained  through QG construction rules for (twice) the conformal weight in $\mathbb C$
 of a random star  made precisely of this effective number $L_{\{h\}}$ of SLEs, while (\ref{QQG0}) is (twice) the
 weight of the original $n$-SLE star without the auxiliary harmonic Brownian paths. (\ref{QQG})  is therefore also exactly
 (twice) the conformal weight $h_{\alpha}=\alpha(\alpha-2\alpha_0)$ corresponding to the holomorphic charge
$\alpha=\alpha_{0,L_{\{h\}}/2}$ of the curve-creating operator $\Psi_{0,L_{\{h\}}/2}$, which gives the final and
rather elegant CG formula for (\ref{xp0}):
\begin{eqnarray}
x(\{h_i\}; 0)+\sum_i h_i=2\alpha_{0,L_{\{h\}}/2}(\alpha_{0,L_{\{h\}}/2}-2\alpha_0)
-2\alpha_{0,n/2}(\alpha_{0,n/2}-2\alpha_0).
\end{eqnarray}

 \subsubsection*{Conclusion} Results (\ref{QG})--(\ref{QQG0}) can be obtained
 by quantum gravity construction rules almost immediately, and have a natural interpretation in that formalism.
 As we saw, they can also be recovered under the form (\ref{xp}) in the fully developed Coulomb gas and conformal field theory approach,
 but though the latter is quite interesting, it also
  appears to be significantly more combersome.

  Its most important aspect is probably the renewed suggestion that a systematic yet rigorous representation
  of  the stochastic properties of SLEs via a Coulomb gas driven by a Gaussian free field must exist \cite{OS}, mimicking the
(heuristic) physics CG approach's  long-established predictive power.
Let us finally mention that the mixed
multifractal moments, introduced here to study the mixed harmonic measure-rotation spectrum of critical random paths,
are amenable to an approach using only the Stochastic Loewner Evolution, that should allow one to establish
rigorously the present results \cite{IABD}.\\

 While writing up this manuscript, we learned
of related work by A. Belikov, I. A. Gruzberg and I. Rushkin.\\

{\bf Acknowledgements:} It is a pleasure to thank
Vladimir Dotsenko for generously sharing his scholarly knowledge of the Coulomb
gas representation, David Kosower and Vincent Pasquier
for their interest and a critical reading of the manuscript, and Emmanuel Guitter for help with the figures.
We also thank the Institute for Pure and Applied Mathematics at UCLA, the IAS/Park City Mathematics Institute and
BD thanks the School of Mathematics of the
Institute for Advanced Study, for their graceful hospitality in successive stays during which this work was done.
BD wishes especially to thank Michael Aizenman and Tom Spencer for their generous hospitality in Princeton.

\end{document}